\documentclass[cernpreprint,  atlasdraft=false, texlive=2025, USenglish, texmf, orcidlogo]{atlasdoc}
\usepackage{mathrsfs}
\usepackage[subfigure=true]{atlaspackage}
\usepackage{atlasbiblatex}

\usepackage{atlasphysics}
\graphicspath{{logos/}{figures/}}

\usepackage{ANA-HIGG-2020-08-PAPER-defs}

\AtlasTitle{Test of \textit{CP} Invariance in Higgs Boson Vector-Boson-Fusion Production Using the $H\to\gamma\gamma$ Channel with the ATLAS Detector}

\AtlasAbstract{%
A test of \textit{CP} invariance in Higgs boson production via vector-boson fusion has been performed in
the $H\rightarrow\gamma\gamma$ channel using 139~fb$^{-1}$ of proton--proton collision data at $\sqrt{s}=13\,\mathrm{TeV}$
collected by the ATLAS detector at the LHC. The Optimal Observable method is used to probe the \textit{CP} structure of interactions
between the Higgs boson and electroweak gauge bosons, as described by an effective field theory.
No sign of \textit{CP} violation is observed in data. Constraints are set on the parameters describing the strength of the \textit{CP}-odd
component in the coupling between the Higgs boson and the electroweak gauge bosons in two effective field theory bases:
$\tilde{d}$ in the HISZ basis and $c_{H\tilde{W}}$ in the Warsaw basis. The results presented are
the most stringent constraints on \textit{CP} violation in the coupling between Higgs and weak bosons.
The 95\% CL constraint on $\tilde{d}$ is derived for the first time and the 95\% CL constraint on
$c_{H\tilde{W}}$ has been improved by a factor of 5 compared to the previous measurement.
}

\AtlasRefCode{HIGG-2020-08}

\PreprintIdNumber{CERN-EP-2022-134}

\AtlasJournal{Phys.\ Rev.\ Lett.}
\AtlasJournalRef{Phys. Rev. Lett. 131 (2023) 061802}
\AtlasDOI{10.1103/PhysRevLett.131.061802}
\hypersetup{pdftitle={ATLAS document},pdfauthor={The ATLAS Collaboration}}

\begin{document}

\maketitle

%

%
%
%
%
%
%
%

%
%
%

The violation of the charge-conjugation and parity (\textit{CP}) symmetry is one of the three Sakharov conditions~\cite{Sakharov:1967dj}
needed to explain the observed baryon asymmetry of the universe.
The only established \textit{CP} violation source is the complex phase in the quark mixing matrix~\cite{Kobayashi:1973fv},
from which the derived magnitude of \textit{CP} violation in the early universe is insufficient to explain the observed value
of the baryon asymmetry~\cite{Gavela:1993ts,Cohen:1993nk,Huet:1994jb}. The discovery of the Higgs boson by the ATLAS
and the CMS experiments~\cite{HIGG-2012-27,CMS-HIG-12-028} at the Large Hadron Collider (LHC)~\cite{LHC} opened
a new direction to search for sources of \textit{CP} violation: the interactions of the Higgs boson.
The Standard Model (SM) Higgs boson ($H$) is even under simultaneous charge-conjugation and parity inversion.
However, \textit{CP} violating interactions are still allowed experimentally.
Any deviation from a pure \textit{CP}-even interaction of the Higgs boson with other SM particles could be a new source of \textit{CP} violation
and also a direct indication of physics beyond the SM (BSM).
The \textit{CP} structure of Higgs boson couplings to electroweak gauge bosons and fermions has been studied extensively by the ATLAS and
the CMS experiments~\cite{HIGG-2018-14,HIGG-2019-01,HIGG-2017-13,HIGG-2019-13,ATLAS:2020rej,CMS-HIG-17-034,CMS-HIG-19-013,CMS-HIG-19-009,CMS-HIG-20-006,CMS-HIG-20-007}.
The results are consistent with the SM prediction, and no sign of \textit{CP} violation has been found yet.

A \textit{CP}-odd component in the Higgs boson coupling to electroweak bosons ($HVV, V=W/Z$) can be described by adding dimension-6 operators
to the SM Lagrangian, using an effective field theory (EFT) approach.
The total matrix element ($\mathscr{M}$) can be written as
\begin{equation}\label{eq:ME_bsm}
\begin{aligned}
|\mathscr{M}|^2 =
&|\mathscr{M}_{\mathrm{SM}}|^2 + 2 c_i \mathrm{Re}(\mathscr{M}^{\ast}_{\mathrm{SM}} \mathscr{M}_{CP\textrm{-odd}}) \\
& + c^2_i |\mathscr{M}_{CP\textrm{-odd}}|^2 .
\end{aligned}
\end{equation}
The first term describes the SM contribution. The second term (interference term) is \textit{CP}-odd,
representing a new source of \textit{CP} violation in Higgs boson couplings, and is parameterized by the Wilson coefficient $c_i$.
The third term (quadratic term) describes a \textit{CP}-even BSM contribution parameterized by $c_i^2 $.
The interference term only affects \textit{CP}-odd observables and does not contribute to \textit{CP}-even observables,
e.g.\ the inclusive cross-section~\cite{Azatov:2016sqh}.

Several methods were developed to construct \textit{CP}-odd observables that can distinguish \textit{CP} violation contributions,
e.g.\ in Refs.~\cite{HIGG-2019-13,CMS-HIG-20-006}.
This study adopts the Optimal Observable~\cite{ALEPH:2004klc,OPAL:2000wbs,Davier:1992nw, Atwood:1991ka, Diehl:1993br} defined as
\begin{linenomath}\begin{equation*}
\OO =2 \mathrm{Re(\mathscr{M}^{*}_{\mathrm{SM}}  \mathscr{M}_{CP\mathrm{-odd}} ) } / {|\mathscr{M}_{\mathrm{SM} } |^{2} }
\end{equation*}\end{linenomath}
to test the \textit{CP} structure of the Higgs boson coupling to electroweak bosons in vector-boson-fusion (VBF) production
and combines event-based information from a multidimensional phase space into a single \textit{CP}-sensitive observable.

The Optimal Observable is evaluated with the momentum fraction $x_{1}$ ($x_{2}$) of the initial-state parton from the proton
moving in the positive (negative) $z$-direction (along the beam), and from the four-momenta of the Higgs boson and two VBF jets.
At the reconstruction level, the momentum fractions are derived as $x_{1,2}^{\mathrm{reco}}=(m_{Hjj}\mathrm{e}^{\pm y_{Hjj}})/\sqrt{s}$
by exploiting energy and momentum conservations of the Higgs boson,
which is built from the two selected photons, and the selected VBF jets.
Here, $m_{Hjj}$ ($y_{Hjj}$) is the invariant mass (rapidity) of the Higgs boson and VBF jet system,
and $\sqrt{s}$ represents the center-of-mass energy of the proton-proton collision.
A detail description of the Optimal Observable calculation can be found in Ref.~\cite{HIGG-2018-14}.

In the SM, the \OO{} distribution is expected to be symmetric with a mean value of zero, and any asymmetrical effects
would indicate contributions from the \textit{CP} violation term, in the absence of rescattering by new light particles
in loops~\cite{Brehmer:2017lrt}.
For a given event, the matrix elements in the \OO{} definition are calculated using
the four-momenta of the Higgs boson and the two forward VBF jets, and have no dependence on the decay mode of the Higgs boson.
This method was first introduced in the $H \rightarrow\tau\tau$ analysis~\cite{HIGG-2018-14} by ATLAS and can be used
in all Higgs boson decay channels.

This Letter reports an analysis to test the \textit{CP} invariance of the $HVV$ coupling by using the Optimal Observable method
in the VBF $H\to\gamma\gamma$ channel, using the 139~\ifb{} of proton--proton ($pp$) collision data at $\sqrt{s}=13~\TeV$
recorded during 2015--2018 with the ATLAS detector.
The VBF signal yield in \OO{} bins is extracted from a simultaneous fit to the diphoton invariant mass spectra split
into the \OO{} bins, which is then used to determine the \textit{CP} violation contributions to the $HVV$ coupling.

Results are interpreted in two EFT bases: the HISZ~\cite{Hagiwara:1993ck} and
Warsaw \cite{Brivio:2017btx, Brivio:2020onw,Grzadkowski:2010es} bases.
The HISZ basis is used in order to combine the results with the previous measurement
from the $H\to\tau\tau$ channel~\cite{HIGG-2018-14}, whereas the Warsaw basis is used to provide measurements
for future combinations with other Higgs boson measurements.
In both bases, three Wilson coefficients multiplying \textit{CP}-odd operators describe possible \textit{CP}-odd couplings between the Higgs boson
and electroweak gauge bosons. In the HISZ basis, $\tilde{d}$ is constrained by assuming $\tilde{d}=\tilde{d}_B$
and setting the third coefficient to zero, as in Ref.~\cite{HIGG-2018-14}. In the Warsaw basis, $c_{H\tilde{W}}$
is constrained by setting $c_{H\tilde{B}}$ and $c_{H\tilde{W}B}$ to zero. In both bases, all \textit{CP}-even operators coefficients are set to zero.
Constraints on all three coefficients in the Warsaw basis were obtained previously in the $H\rightarrow ZZ$ channel~\cite{ATLAS:2020rej, CMS-HIG-19-009}
and $H\rightarrow\gamma\gamma$ channel using differential cross-sections~\cite{HIGG-2019-13}.
The measurements have significant correlations since these channels cannot distinguish between
the three operators. The VBF topology in this analysis is mainly sensitive to $c_{H\tilde{W}}$
and could help to reduce this correlation.

%


%

%
%
%

The ATLAS detector~\cite{PERF-2007-01, PIX-2018-001, ATLAS-TDR-19} is a multipurpose particle detector
with a forward-backward symmetric cylindrical geometry and near 4$\pi$ coverage in solid angle \cite{atlasdetector}.
The trigger system consists of a hardware-based first-level trigger and a software-based high-level trigger~\cite{TRIG-2016-01}.
Events used in this analysis were accepted by a diphoton trigger requiring the leading and subleading photons to have
transverse energies ($E_{\mathrm{T}}$) greater than 35~\GeV\ and 25~\GeV, respectively, during the whole data-taking period.
This trigger had a \textit{Loose} photon identification requirement in 2015--2016~\cite{TRIG-2018-05}, but due to the increasing
instantaneous luminosity the identification requirement was tightened for data-taking in 2017--2018~\cite{TRIG-2018-05}.
In addition, a single-photon trigger with \textit{Loose} identification criteria and an $E_{\mathrm{T}}$ threshold
of 120 (140)~\GeV\ in 2015--2016 (2017--2018) was used to recover events with collimated diphoton pairs with very high
transverse momentum (\pt)~\cite{TRIG-2018-05}. The average trigger efficiency is over 98\% for events passing the full diphoton
event selection for this analysis~\cite{TRIG-2018-05}. An extensive software suite~\cite{ATL-SOFT-PUB-2021-001} is used in
the reconstruction and analysis of real and simulated data, in detector operations, and in the trigger
and data acquisition systems of the experiment.

Higgs boson production via VBF was simulated with \POWHEGBOX[v2]~\cite{Nason:2009ai} using the \PDFforLHC[15nlo]~\cite{Butterworth:2015oua} parton distribution function (PDF) set.
The generation is accurate to next-to-leading-order (NLO) in QCD,
and the total cross-section is normalized to a calculation including QCD corrections at full NLO and approximate next-to-next-to-leading-order (NNLO) accuracy as well as electroweak (EW) corrections at full NLO accuracy~\cite{Ciccolini:2007jr,Ciccolini:2007ec,Bolzoni:2010xr}.
Higgs boson production via gluon-gluon fusion (ggF) was modeled at NNLO accuracy in QCD using \POWHEGBOX[v2]~\cite{Hamilton:2013fea,Hamilton:2015nsa} and the NNLO family of \PDFforLHC[15] PDFs.
The simulation achieves NNLO accuracy for arbitrary inclusive \(gg\to H\) observables by reweighting the Higgs boson
rapidity spectrum in \textsc{Hj}-\MINLO~\cite{Hamilton:2012np,Campbell:2012am,Hamilton:2012rf} to that in HNNLO~\cite{Catani:2007vq},
and the total cross-section is normalized to a prediction calculated at next-to-next-to-next-to-leading-order (N\({}^3\)LO) accuracy
in QCD and including NLO EW corrections~\cite{deFlorian:2016spz,Anastasiou:2016cez,Anastasiou:2015ema,Dulat:2018rbf,Harlander:2009mq,Harlander:2009bw,Harlander:2009my,Pak:2009dg,Actis:2008ug,Actis:2008ts,Bonetti:2018ukf,Bonetti:2018ukf}.
Other Higgs boson production processes, e.g.\ in association with a vector boson ($VH$) or top quark(s) ($ttH$, $tH$),
were also modeled using \POWHEGBOX[v2].
Prompt diphoton production ($\gamma\gamma$) was simulated with the \SHERPA[2.2.4]~\cite{Sherpa:2019gpd} generator.
More details can be found in Ref.~\cite{HIGG-2019-13}.

To simulate the effects of nonzero values of $\tilde{d}$ and $c_{H\tilde{W}}$ in the $HVV$ vertex, a reweighting
method is implemented for the HISZ basis and Warsaw basis, respectively, and applied to the aforementioned
SM VBF signal sample.
For the $\tilde{d}$ coefficient in the HISZ basis, as detailed in Ref.~\cite{HIGG-2018-14}, two weights are calculated
by the HAWK program~\cite{Ciccolini:2007jr, Ciccolini:2007ec, Denner:2014cla} for each event
using generator-level information:
$w_1=2\mathrm{Re(\mathscr{M}^{*}_{\mathrm{SM}} \mathscr{M}_{CP\textrm{-odd}} ) } / {|\mathscr{M}_{\mathrm{SM} } |^{2} } $
and
$w_2=\mathrm{|\mathscr{M}_{CP\textrm{-odd}} |^2} / {|\mathscr{M}_{\mathrm{SM} } |^{2} } $
with a specific amount of \textit{CP} mixing (given in terms of $\tilde{d}$), to model the contribution from the interference term
and the quadratic term, respectively, as shown in Eq.~(\ref{eq:ME_bsm}). For the interpretation in the Warsaw basis,
a reweighting of the reconstructed \OO{} distribution at different values of $c_{H\tilde{W}}$ is obtained by
\begin{equation*}
\frac{{d} \sigma}{{d} \OO} = \left( \frac{{d}\sigma}{{d} \OO} \right)^{\mathrm{NLO}} \times \left( \frac{{d}\sigma}{{d} \OO} \right)_{c_{H\tilde{W}}}^{\mathrm{MG5} } / \left( \frac{{d}\sigma}{{d} \OO} \right)_{c_{H\tilde{W}}=0}^{\mathrm{MG5}},
\end{equation*}
where `MG5' labels the prediction from \MADGRAPH~\cite{Alwall:2014hca,Frederix:2012ps} using
SMEFTSim~\cite{Brivio:2017btx, Brivio:2020onw},
and `NLO' labels the aforementioned SM VBF signal sample. \MADGRAPH events for nonzero values of $c_{H\tilde{W}}$
were generated setting the scale of new physics $\Lambda=1~\TeV$ and fixing all other Wilson coefficients to zero.
For both interpretations, higher-order QCD and electroweak corrections are assumed to factorize from the new-physics effects.
Limits in the two bases are extracted from the effect of the interference-only term and also from the effect of
the interference-plus-quadratic terms.
The \OO{} value is calculated using HAWK because the $HWW$ operators in the two EFT bases are similar.
HAWK uses the HISZ basis assuming $\tilde{d}=\tilde{d}_B$, which corresponds to $c_{H\tilde{W}}=c_{H\tilde{B}}$
for the Warsaw basis. However, since $c_{H\tilde{B}}$ has negligible impact on VBF, only $c_{H\tilde{W}}$
is varied (setting $c_{H\tilde{B}}=0$) and the computed \OO{} is assumed to be equally optimal for $c_{H\tilde{W}}$ only.

All generated events were passed through a full simulation of the ATLAS detector response~\cite{SOFT-2010-01}
using \GEANT~\cite{GEANT4:2002zbu}, except the \SHERPA $\gamma\gamma$ sample, which was passed through a fast parametric
simulation of the detector response~\cite{SOFT-2010-01}. The effects of multiple $pp$ interactions in the same
or neighboring bunch crossings (pileup) are included by overlaying events generated with \PYTHIA[8]~\cite{Sjostrand:2014zea}.
Events are weighted such that the distribution of the average number of interactions per bunch crossing matches that observed in data.
%


%

%
%
%

Photons are reconstructed from variable-size topological clusters formed from electromagnetic calorimeter cells with
significant energy deposits and from tracks, initiated by converted photons, measured in the inner detector (ID)~\cite{EGAM-2018-01}.
Events must have at least two photon candidates outside the calorimeter's transition region between the barrel and the end-cap, $1.37<|\eta|<1.52$,
and within $|\eta|<2.37$, where the two leading (highest-\ET) photons are used to reconstruct the Higgs boson candidate and the primary vertex of
the event~\cite{HIGG-2013-08}. The diphoton invariant mass $m_{\gamma\gamma}$ is required to be in the range 105--160~\GeV.
The leading and subleading photons are further required to have $E_{\mathrm{T}}/m_{\gamma\gamma}$ greater than 0.35 and 0.25,
respectively, and fulfill the \textit{Tight} identification selection and \textit{Tight} calorimetric
and track-based isolation requirement~\cite{EGAM-2018-01}. Jets are reconstructed using the anti-\kt{}
algorithm~\cite{Cacciari:2008gp, Cacciari:2011ma} with a radius parameter $R=0.4$ from inputs formed
with a particle-flow algorithm~\cite{JETM-2018-06}, which uses information from both the calorimeter and the ID.
Jet candidates are required to have $p_{\mathrm{T}}>30\,\mathrm{GeV}$ and $|\eta|<4.4$. To suppress
jets from pileup collisions, jet candidates with $|\eta|<2.4$ and $p_{\mathrm{T}}<60\,\mathrm{GeV}$ are required
to pass the \textit{Tight} jet vertex tagger (JVT) selection~\cite{PERF-2014-03}. For jets with $|\eta|\ge 2.4$,
the \textit{Loose} forward JVT selection~\cite{PERF-2016-06} is applied to remove pileup jet contamination.
To construct the region enriched with VBF signal events, two loose criteria are applied: events must have at least two jets
with pseudorapidity separation $|\Delta \eta_{jj}|>2$ and Zeppenfeld variable~\cite{Andersen:2010zx}
$\eta^{\mathrm{Zepp}}=|\eta_{\gamma\gamma}-(\eta_{j1}+\eta_{j2})/2| < 5$.

To increase the VBF signal purity, two boosted decision trees (BDT)~\cite{TMVA} are trained. $\mathrm{BDT_{VBF/ggF}}$
is used to separate VBF signal from ggF events, which are the major background from Higgs boson production.
$\mathrm{BDT_{VBF/Continuum}}$ is used to distinguish VBF $H\rightarrow\gamma\gamma$ events from continuum background events,
which consist of the prompt diphoton events ($\gamma\gamma$) and events where one or two of the photon candidates originate
from jets misidentified as photons ($\gamma j$ or $jj$). The $\gamma\gamma$ events, which are the dominant component
of the continuum background, are obtained from simulation, while $\gamma j$ and $jj$ events are obtained from
dedicated data control regions, as described later. The two BDTs use the same input variables:
invariant mass of the dijet system formed by the two leading jets ($m_{jj}$),
pseudorapidity separation of the dijet system ($\Delta \eta_{jj}$),
$\pt$ of the Higgs boson and the leading two jets ($p_{\mathrm{T}}^{Hjj}$),
azimuthal angle between the diphoton and dijet systems ($\Delta \phi (\gamma\gamma, jj)$),
minimum angular separation between the photons and the two leading jets ($\Delta R^{\mathrm{min}}_{\gamma j}$),
$\eta^{\mathrm{Zepp}}$ and
perpendicular projection of the diphoton $\pt$ onto the diphoton thrust axis ($p^{\gamma\gamma}_{\mathrm{Tt}}$)~\cite{HIGG-2013-02}.
These input variables are all \textit{CP}-even
to be insensitive to the \textit{CP} property of the VBF signal and to have negligible correlation with $m_{\gamma\gamma}$.
Figure~\ref{fig:VBFBDT} shows the BDT output distributions of the VBF signal, the ggF background, the continuum background,
and the data in the $m_{\gamma\gamma}$ sideband ($m_{\gamma\gamma}\in[105, 118]~\GeV$ or $[132, 160]~\GeV$).
The comparison between the continuum background and the sideband data shows the
continuum background used in the BDT training is well modeled.
Events are categorized as follows:
firstly, a requirement is placed on $\mathrm{BDT_{VBF/ggF}}$ to separate events into `tight' (T) and `loose' (L) regions.
The ratio of VBF signal to ggF background is improved by a factor of ten in the `tight' region.
Then, two independent requirements on $\mathrm{BDT_{VBF/Continuum}}$ are applied to the `tight' and `loose' regions to maximize
the combined significance of the VBF signal. Three signal regions are defined: TT, TL, and LT,
where the first (second) letter corresponds to the $\mathrm{BDT_{VBF/ggF}}$ ($\mathrm{BDT_{VBF/Continuum}}$) separation type.
More details on the BDT input variables and the categorization requirements can be found in the Supplemental Material~\cite{SupplementalMaterial}.
In the TT and TL categories, the dominant Higgs boson backgrounds are from the ggF process, and the contributions
from non-ggF Higgs processes, e.g. $VH$, $ttH$ and $tH$, are found to be negligible. In the LT category, Higgs boson backgrounds are
still mostly from the ggF process, while those from non-ggF Higgs processes increase to about 1\%--3\% of the VBF event yield.
This novel BDT-based strategy improves the significance of VBF signal by 10\% with respect to
the latest $H\rightarrow\gamma\gamma$ analyses \cite{HIGG-2020-16} with the same dataset.

\begin{figure}[thbp]
\centering
\includegraphics[width=0.9\columnwidth]{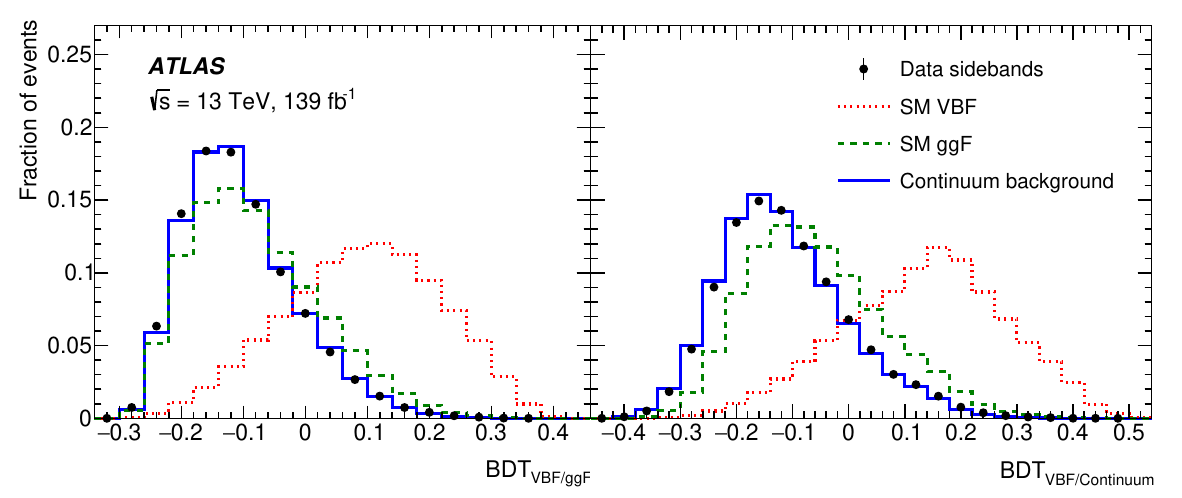}
\caption{Distribution of the output of $\mathrm{BDT_{VBF/ggF}}$ (left) and $\mathrm{BDT_{VBF/continuum}}$ (right).
The comparison between the continuum background and the sideband data indicates the
continuum background used in the BDT training is well modeled.
}
\label{fig:VBFBDT}
\end{figure}


%

%
%
%

%
%
%
The signal yield is extracted via a combined unbinned maximum-likelihood estimation
applied to the $m_{\gamma\gamma}$ distribution of observed data in each \OO{} bin, as shown in Figure~\ref{fig:MggOO}.
Both the signal and background shapes are modeled with analytic functions.
The $H\to\gamma\gamma$ signal shape is described by a double-sided Crystal Ball (DSCB) function~\cite{HIGG-2019-13},
consisting of a Gaussian distribution in the region around the peak, continued by power-law tails at lower
and higher $m_{\gamma\gamma}$ values. The parameters of the DSCB function in each category are obtained by a fit to
the simulated VBF sample, as well as other Higgs boson production modes in proportion to their SM cross-sections.
The modeling of the continuum background relies on both simulation and data-driven methods.
The $m_{\gamma\gamma}$ shape of the $\gamma\gamma$ component is estimated using the \SHERPA sample,
while the $m_{\gamma\gamma}$ shapes of the $\gamma j$ and $jj$ components are obtained using data control regions formed
by inverting the \textit{Tight} photon identification and isolation requirements. The template is then built by summing
the $\gamma\gamma$, $\gamma j$ and $jj$ components, where their fractions are measured in data using a two-dimensional
double-sideband method~\cite{STDM-2011-05}.  The composition of the continuum background is found to be approximately
85\% $\gamma\gamma$ events and 15\% $\gamma j+jj$ events. The background templates are smoothed using Gaussian process
regression (GPR)~\cite{Frate:2017mai} with the Gibbs kernel to reduce fluctuations due to the limited sample size.
The $m_{\gamma\gamma}$ distribution of the continuum background is found to have a smoothly falling shape. The analytic function
chosen to model the continuum background is either a power-law function, a Bernstein polynomial~\cite{HIGG-2019-13},
or an exponential function of a polynomial, and it is selected
for each \OO{} bin independently. The selected function should have the smallest spurious signal, defined as the systematic bias
in the fitted signal yield due to differences between the fit function choice and the background template. The coefficients of these functions are considered
to be independent across categories, and in all cases are treated as free parameters in the fits to data.
More details can be found in Ref.~\cite{HIGG-2019-13}.

\begin{figure}[htbp]
\centering
\includegraphics[width=0.8\columnwidth]{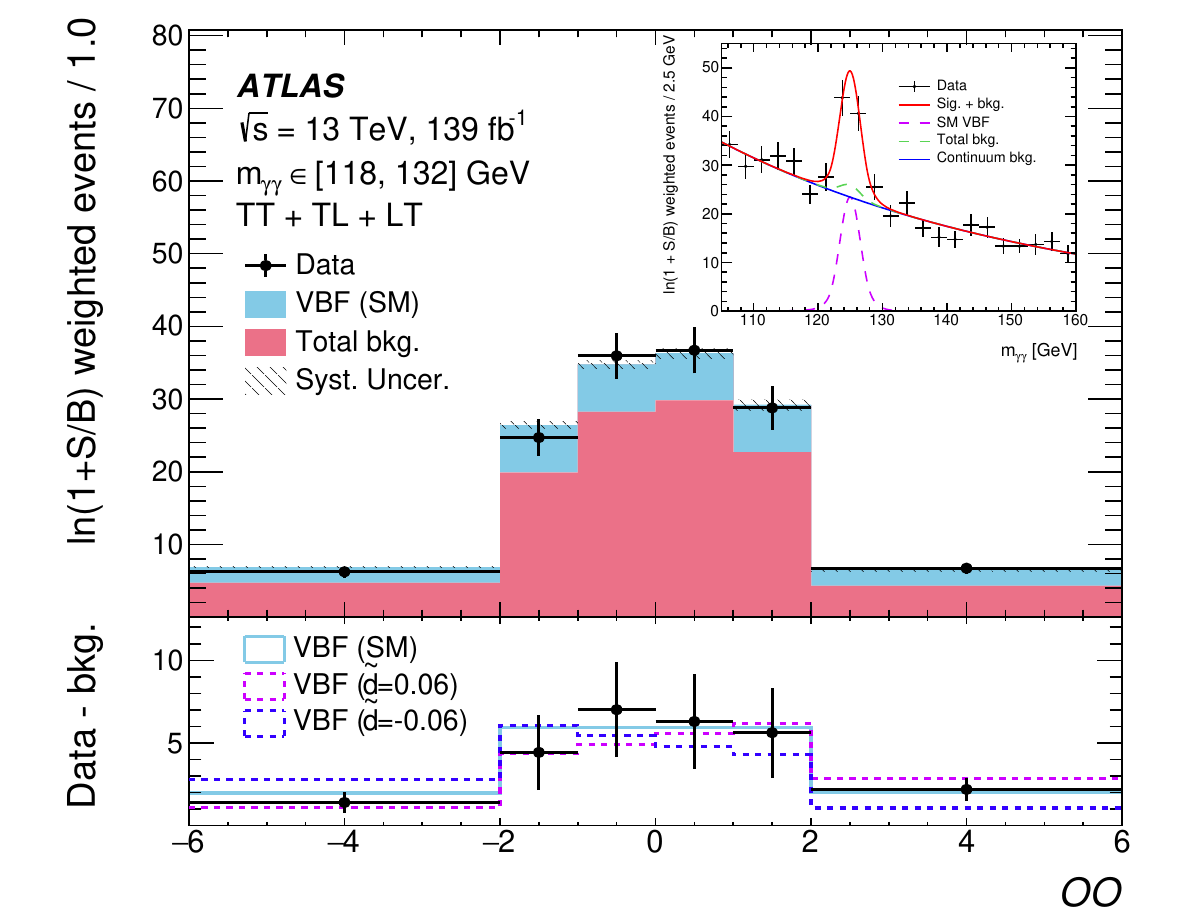}
\caption{
Distribution of the optimal observable \OO{} for events with $m_{\gamma\gamma}\in[118, 132]~\GeV$. Contributions in three signal regions
are summed together with a weight of $\ln(1+S/B)$ for each signal region,
where $S$ and $B$ are the expected yields of signal and background events with $m_{\gamma\gamma}\in[118, 132]~\GeV$.
The overflow and underflow are included in the highest and lowest bin, respectively.
The uncertainty band shown includes all systematic uncertainties.
The weighted summed $m_{\gamma\gamma}$ distribution of data events is shown in the inner panel along with the signal
and background contributions. The lower panel is the \OO{} distribution in data after subtraction of all backgrounds,
in comparison with the SM VBF process, and VBF processes with $\tilde{d}=0.06$ and $\tilde{d}=-0.06$.
The sensitivity to $\tilde{d}$ is dominated by the tails of the \OO{} distribution.
}
\label{fig:MggOO}
\end{figure}


%

%
%
%

An unbinned likelihood is constructed with the $m_{\gamma\gamma}$ spectra of each \OO{} bin in signal regions TT, TL and LT.
The negative log-likelihood ($\mathrm{NLL}$) is evaluated for various $\tilde{d}$ and $c_{H\tilde{W}}$ hypotheses.
Confidence intervals are obtained by reading values off the $\mathrm{NLL}$ curve, which is constructed by interpolating
between the points with spline functions. The normalization of the signal is allowed to float in the fit.
The analysis therefore exploits only the shape of the distribution of the Optimal Observable, and ignores
the potential dependence of the inclusive cross-section on \textit{CP}-mixing scenarios.
If present, any BSM \textit{CP}-even effects would mainly change the normalization, and produce very small symmetric
changes in the \OO{} distribution, which are found to not bias the parameter of interest for the \textit{CP}-odd effect.
All other Higgs boson production modes are considered as backgrounds and are normalized to their SM predicted yields.
The expected $\Delta \mathrm{NLL}$ curve is obtained using a pseudo-dataset where the event yields and distributions
in the signal regions are set to the SM expectations for both the signal and background processes.

Both the theoretical and experimental systematic uncertainties are incorporated into the likelihood model of the measurement
as nuisance parameters. Theoretical uncertainties arise from the modeling of VBF and ggF processes because of the
missing higher-order terms in the perturbative QCD calculations, the modeling of the underlying event and parton shower,
the parton distribution functions, and the value of $\alphas$. These uncertainties are estimated by following the procedure
described in Ref.~\cite{HIGG-2019-13}. The experimental uncertainties include the uncertainties in the photon energy scale
and resolution~\cite{EGAM-2018-01}, the jet energy scale and resolution~\cite{JETM-2018-05}, the luminosity measurement,
and the modeling of pileup events and the photon identification and isolation criteria~\cite{PERF-2017-02}.
The spurious signal that could arise from mismodeling of continuum background is estimated in each \OO{} bin.

Figure~\ref{fig:NLL} shows the $\Delta \mathrm{NLL}$ curves as functions of $\tilde{d}$ or $c_{H\tilde{W}}$.
Here, the $\tilde{d}$ results use the interference-plus-quadratic terms in Eq.~(\ref{eq:ME_bsm}), while the $c_{H\tilde{W}}$
results use only the interference term.  The confidence intervals for the two scenarios, interference-only and
interference-plus-quadratic, are shown in Table~\ref{tab:results}. The difference between the results in the two scenarios is found to be small.
The results are compatible with the SM and the precision is limited by the statistical uncertainty of the data.
For example, the total impact on the 95\% confidence intervals of $\tilde{d}$ from the systematic uncertainty is less than 2\%.
The measurement is sensitive enough to determine an observed 95\% confidence interval for $\tilde{d}$,
which was not achieved in previous analyses.
The expected 68\% confidence interval shown for the $H\rightarrow \tau\tau$ channel in Table~\ref{tab:results} differs slightly
from that presented in Ref.~\cite{HIGG-2018-14}, where the expected $H\rightarrow \tau\tau$ results were obtained
with the nuisance parameters constrained only by the control regions. In the present analysis, the expected results
are obtained with the nuisance parameters constrained by both the control regions and signal regions.
The saturation of the $\Delta \mathrm{NLL}$ shape at larger values of $\tilde{d}$ is a result of the dominance of the quadratic term.
This was a limiting factor in the $H\rightarrow \tau\tau$ analysis where this saturation together with the larger statistical
and systematic uncertainties prevented setting intervals at the 95\% confidence level, the level that is most commonly used to constrain
the corresponding EFT operators.
This letter reports a significantly improved expected 95\% confidence interval compared to previous $H\rightarrow \tau\tau$ analysis and presents the observed 95 \% confidence interval for the first time.
The 95\% confidence interval for $c_{H\tilde{W}}$ obtained using the interference-only term is a factor of five more restrictive than
in the $H\rightarrow\gamma\gamma$ differential measurement reported in Ref.~\cite{HIGG-2019-13} because of the dedicated BDTs
for the VBF signal selection and the use of the Optimal Observable. The 68\% confidence interval for $c_{H\tilde{W}}$
is about twice as restrictive as that from either the ATLAS or CMS $H\rightarrow ZZ$ four-lepton
analysis~\cite{ATLAS:2020rej, CMS-HIG-19-009}.
The luminosity uncertainty of the data in 2015--2016, the uncertainties of the photon energy scale and resolution,
and the theoretical uncertainty of VBF and ggF processes are correlated in the combination. The jet-related uncertainties
are not correlated since a different jet reconstruction technique was used in the $H\rightarrow \tau\tau$ analysis.

\begin{figure}[t]
\centering
\subfigure[]{\includegraphics[width=0.48\columnwidth]{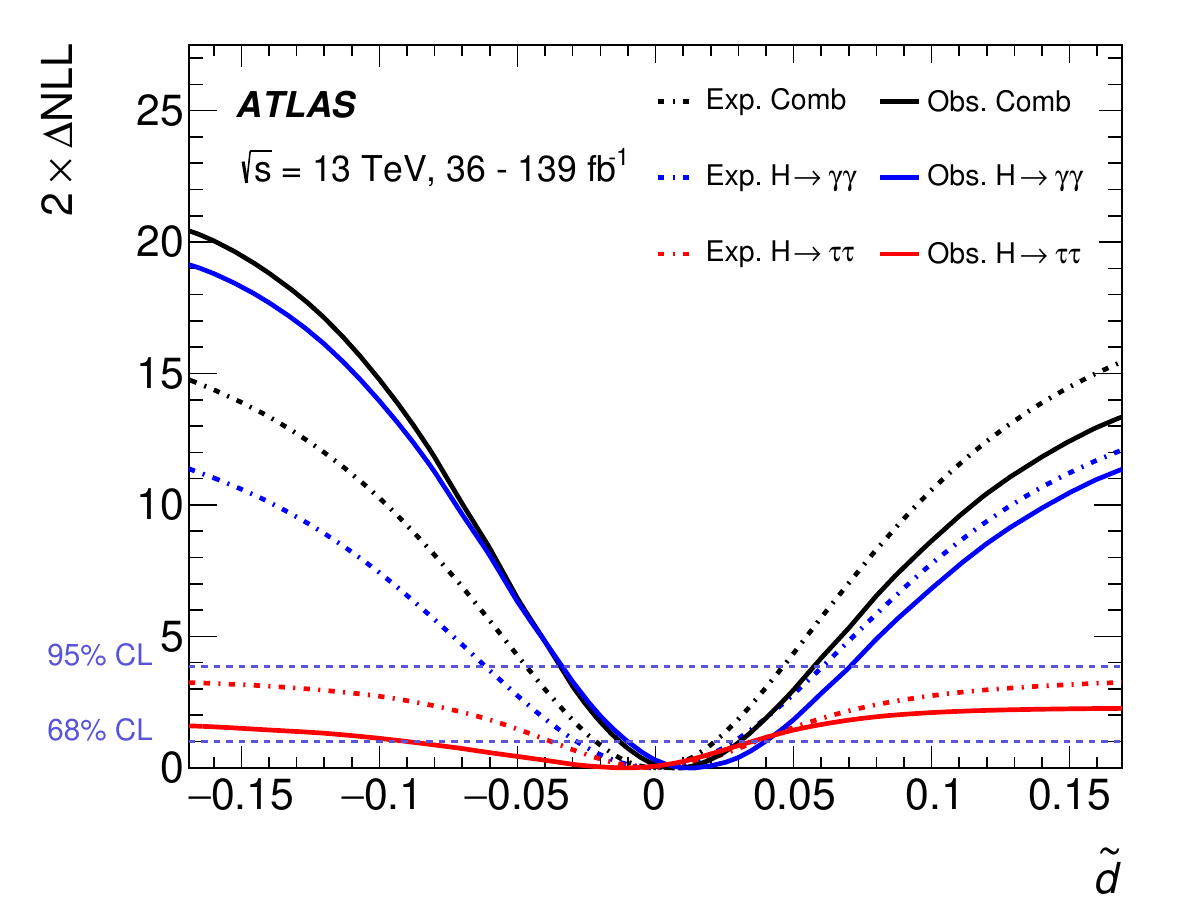}}
\subfigure[]{\includegraphics[width=0.48\columnwidth]{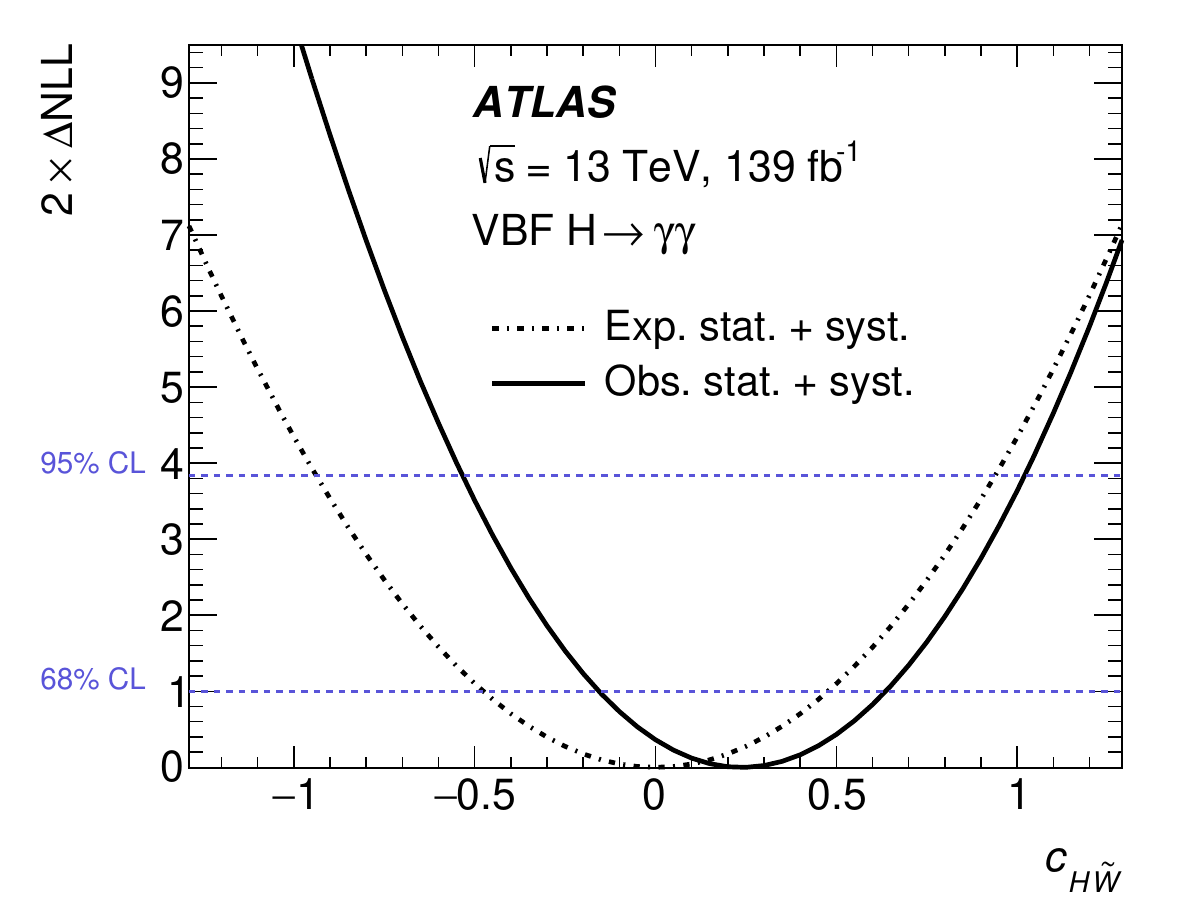}}
\caption{
$\Delta\mathrm{NLL}$ curves as a function of (a) $\tilde{d}$ and (b) $c_{H\tilde{W}}$.
In figure (a), the $\Delta\mathrm{NLL}$ of $\tilde{d}$ considers the interference-plus-quadratic terms,
whereas in figure (b) the $\Delta\mathrm{NLL}$ of $c_{H\tilde{W}}$ considers the interference-only term.
The solid lines are the observed results, while the dashed lines are the expected results.
In figure (a), the blue lines represent the results of this analysis, while the red lines represent the results
from the $H\rightarrow \tau\tau$ analysis \cite{HIGG-2018-14}. The black lines show the combination of these two analyses.
For all figures, the dashed horizontal lines show the values of $\Delta\mathrm{NLL}$ used to define the 68\% and 95\% confidence intervals.
}
\label{fig:NLL}
\end{figure}

\begin{table*}[thbp]
\centering
\caption{
Observed (noted as `obs.') and expected (noted as `exp.') 68\% and 95\% confidence intervals for $\tilde{d}$ and $c_{H\tilde{W}}$. Results for scenarios
with the interference-only (noted as `inter.\ only') term and interference-plus-quadratic terms (noted as `inter.+quad.')
are both presented.
Combined results for $\tilde{d}$ including the  $H\rightarrow\tau\tau$ analysis~\cite{HIGG-2018-14} are shown.
The expected results of $H\rightarrow\tau\tau$ are slightly different from Ref.~\cite{HIGG-2018-14}
due to the different correlation scheme between their signal region and control region.
}
\renewcommand{\arraystretch}{1.2}
\begin{tabular}{l|c|c|c|c}
\hline
\hline
& 68\% (exp.) & 95\% (exp.)  & 68\% (obs.) & 95\% (obs.) \\
\hline
$\tilde{d}$~(inter.\ only)  & [$-$0.027, 0.027] & [$-$0.055, 0.055] & [$-$0.011, 0.036]  & [$-$0.032, 0.059]  \\ \hline
$\tilde{d}$~(inter.+quad.) & [$-$0.028, 0.028] & [$-$0.061, 0.060] & [$-$0.010, 0.040]  & [$-$0.034, 0.071]  \\ \hline
$\tilde{d}$ from $H\rightarrow\tau\tau$  & [$-$0.038, 0.036] & $\cdots$ & [$-$0.090, 0.035]  & $\cdots$  \\ \hline

Combined $\tilde{d}$  & [$-$0.022, 0.021] & [$-$0.046, 0.045] & [$-$0.012, 0.030] & [$-$0.034, 0.057]  \\ \hline\hline
$c_{H\tilde{W}}$~(inter.\ only) & [$-$0.48, 0.48] & [$-$0.94, 0.94] & [$-$0.16, 0.64] & [$-$0.53, 1.02]  \\ \hline
$c_{H\tilde{W}}$~(inter.+quad.) & [$-$0.48, 0.48] & [$-$0.95, 0.95] & [$-$0.15, 0.67] & [$-$0.55, 1.07]  \\
\hline
\hline
\end{tabular}
\label{tab:results}
\end{table*}


%

\FloatBarrier

In conclusion, a test of \textit{CP} invariance in Higgs boson production via vector-boson fusion is performed in
the $H\rightarrow\gamma\gamma$ channel using 139~\ifb{} of $\sqrt{s}=13\,\TeV$ proton--proton collision data collected
by the ATLAS detector at the LHC. The Optimal Observable method is used to probe \textit{CP}-violating interactions between the Higgs boson
and electroweak gauge bosons described by an effective field theory.
The results are compatible with the SM. No sign of \textit{CP} violation is observed in the Optimal Observable distributions.
The constraints on \textit{CP}-violating effects in the $HVV$ coupling are the most stringent to date. They allow
68\% and 95\% confidence intervals to be set for parameters describing the strength of the \textit{CP}-odd component in
the $HVV$ coupling in two effective field theory bases: $\tilde{d}$ in the HISZ basis and $c_{H\tilde{W}}$ in the Warsaw basis.
The sensitivity is sufficient to set a 95\% confidence interval for $\tilde{d}$ for the first time, and the constraints on $\tilde{d}$
are tightened further by combining them with previous results from the $H\rightarrow\tau\tau$ channel.
The constraints on $c_{H\tilde{W}}$ are
about twice as restrictive as those from either the ATLAS or CMS four-lepton analysis.


%
%
\section*{Acknowledgments}
%

%
%

%
%

We thank CERN for the very successful operation of the LHC, as well as the
support staff from our institutions without whom ATLAS could not be
operated efficiently.

We acknowledge the support of
ANPCyT, Argentina;
YerPhI, Armenia;
ARC, Australia;
BMWFW and FWF, Austria;
ANAS, Azerbaijan;
CNPq and FAPESP, Brazil;
NSERC, NRC and CFI, Canada;
CERN;
ANID, Chile;
CAS, MOST and NSFC, China;
Minciencias, Colombia;
MEYS CR, Czech Republic;
DNRF and DNSRC, Denmark;
IN2P3-CNRS and CEA-DRF/IRFU, France;
SRNSFG, Georgia;
BMBF, HGF and MPG, Germany;
GSRI, Greece;
RGC and Hong Kong SAR, China;
ISF and Benoziyo Center, Israel;
INFN, Italy;
MEXT and JSPS, Japan;
CNRST, Morocco;
NWO, Netherlands;
RCN, Norway;
MEiN, Poland;
FCT, Portugal;
MNE/IFA, Romania;
MESTD, Serbia;
MSSR, Slovakia;
ARRS and MIZ\v{S}, Slovenia;
DSI/NRF, South Africa;
MICINN, Spain;
SRC and Wallenberg Foundation, Sweden;
SERI, SNSF and Cantons of Bern and Geneva, Switzerland;
MOST, Taiwan;
TENMAK, T\"urkiye;
STFC, United Kingdom;
DOE and NSF, United States of America.
In addition, individual groups and members have received support from
BCKDF, CANARIE, Compute Canada and CRC, Canada;
PRIMUS 21/SCI/017 and UNCE SCI/013, Czech Republic;
COST, ERC, ERDF, Horizon 2020 and Marie Sk{\l}odowska-Curie Actions, European Union;
Investissements d'Avenir Labex, Investissements d'Avenir Idex and ANR, France;
DFG and AvH Foundation, Germany;
Herakleitos, Thales and Aristeia programmes co-financed by EU-ESF and the Greek NSRF, Greece;
BSF-NSF and MINERVA, Israel;
Norwegian Financial Mechanism 2014-2021, Norway;
NCN and NAWA, Poland;
La Caixa Banking Foundation, CERCA Programme Generalitat de Catalunya and PROMETEO and GenT Programmes Generalitat Valenciana, Spain;
G\"{o}ran Gustafssons Stiftelse, Sweden;
The Royal Society and Leverhulme Trust, United Kingdom.

The crucial computing support from all WLCG partners is acknowledged gratefully, in particular from CERN, the ATLAS Tier-1 facilities at TRIUMF (Canada), NDGF (Denmark, Norway, Sweden), CC-IN2P3 (France), KIT/GridKA (Germany), INFN-CNAF (Italy), NL-T1 (Netherlands), PIC (Spain), ASGC (Taiwan), RAL (UK) and BNL (USA), the Tier-2 facilities worldwide and large non-WLCG resource providers. Major contributors of computing resources are listed in Ref.~\cite{ATL-SOFT-PUB-2021-003}.

%
%


%
%
%
%
%
%
%

%
%
%
%
%
%

\clearpage
\printbibliography

@article{Azatov:2016sqh,
    author = "Azatov, Aleksandr and Contino, Roberto and Machado, Camila S. and Riva, Francesco",
    title = "{Helicity selection rules and noninterference for BSM amplitudes}",
    eprint = "1607.05236",
    archivePrefix = "arXiv",
    primaryClass = "hep-ph",
    reportNumber = "CERN-TH-2016-165",
    doi = "10.1103/PhysRevD.95.065014",
    journal = "Phys. Rev. D",
    volume = "95",
    number = "6",
    pages = "065014",
    year = "2017"
}

@Article{Sakharov:1967dj,
  author  = {Sakharov, A. D.},
  journal = {Pisma Zh. Eksp. Teor. Fiz.},
  title   = {{Violation of CP Invariance, C asymmetry, and baryon asymmetry of the universe}},
  year    = {1967},
  pages   = {32--35},
  volume  = {5},
  doi     = {10.1070/PU1991v034n05ABEH002497},
}

@Article{Kobayashi:1973fv,
  author       = {Kobayashi, Makoto and Maskawa, Toshihide},
  journal      = {Prog. Theor. Phys.},
  title        = {{CP Violation in the Renormalizable Theory of Weak Interaction}},
  year         = {1973},
  pages        = {652--657},
  volume       = {49},
  doi          = {10.1143/PTP.49.652},
  reportnumber = {KUNS-242},
}

@Article{Huet:1994jb,
  author        = {Huet, Patrick and Sather, Eric},
  journal       = {Phys. Rev. D},
  title         = {{Electroweak baryogenesis and standard model CP violation}},
  year          = {1995},
  pages         = {379--394},
  volume        = {51},
  archiveprefix = {arXiv},
  doi           = {10.1103/PhysRevD.51.379},
  eprint        = {hep-ph/9404302},
  reportnumber  = {SLAC-PUB-6479},
}

@Article{Gavela:1993ts,
  author        = {Gavela, M. B. and Hernandez, P. and Orloff, J. and Pene, O.},
  journal       = {Mod. Phys. Lett. A},
  title         = {{Standard model CP violation and baryon asymmetry}},
  year          = {1994},
  pages         = {795--810},
  volume        = {9},
  archiveprefix = {arXiv},
  doi           = {10.1142/S0217732394000629},
  eprint        = {hep-ph/9312215},
  reportnumber  = {CERN-TH-7081-93, LPTHE-ORSAY-93-48, HUTP-93-A036, HD-THEP-93-45},
}

@Article{Cohen:1993nk,
  author        = {Cohen, Andrew G. and Kaplan, D. B. and Nelson, A. E.},
  journal       = {Ann. Rev. Nucl. Part. Sci.},
  title         = {{Progress in Electroweak Baryogenesis}},
  year          = {1993},
  pages         = {27--70},
  volume        = {43},
  archiveprefix = {arXiv},
  doi           = {10.1146/annurev.ns.43.120193.000331},
  eprint        = {hep-ph/9302210},
  reportnumber  = {UCSD-PTH-93-02, BUHEP-93-4},
}

@Article{Grzadkowski:2010es,
  author        = {Grzadkowski, B. and Iskrzynski, M. and Misiak, M. and Rosiek, J.},
  journal       = {JHEP},
  title         = {{Dimension-six terms in the Standard Model Lagrangian}},
  year          = {2010},
  pages         = {085},
  volume        = {10},
  archiveprefix = {arXiv},
  doi           = {10.1007/JHEP10(2010)085},
  eprint        = {1008.4884},
  primaryclass  = {hep-ph},
  reportnumber  = {IFT-9-2010, TTP10-35},
}

@article{Frate:2017mai,
      author         = "Frate, Meghan and Cranmer, Kyle and Kalia, Saarik and
                        Vandenberg-Rodes, Alexander and Whiteson, Daniel",
      title          = "{Modeling Smooth Backgrounds and Generic Localized
                        Signals with Gaussian Processes}",
      year           = "2017",
      eprint         = "1709.05681",
      archivePrefix  = "arXiv",
      primaryClass   = "physics.data-an",
      SLACcitation   = "%%CITATION = ARXIV:1709.05681;%%"
}

@Article{GEANT4:2002zbu,
  author        = {{GEANT4 Collaboration}},
  journal       = {Nucl. Instrum. Meth. A},
  title         = {{\textsc{Geant}4--a simulation toolkit}},
  year          = {2003},
  pages         = {250--303},
  volume        = {506},
  collaboration = {GEANT4},
  doi           = {10.1016/S0168-9002(03)01368-8},
  reportnumber  = {SLAC-PUB-9350, FERMILAB-PUB-03-339, CERN-IT-2002-003},
}

@Article{Sherpa:2019gpd,
  author        = {Bothmann, Enrico and others},
  journal       = {SciPost Phys.},
  title         = {{Event Generation with Sherpa 2.2}},
  year          = {2019},
  number        = {3},
  pages         = {034},
  volume        = {7},
  archiveprefix = {arXiv},
  collaboration = {Sherpa},
  doi           = {10.21468/SciPostPhys.7.3.034},
  eprint        = {1905.09127},
  primaryclass  = {hep-ph},
  reportnumber  = {FERMILAB-PUB-19-218-T, SLAC-PUB-17433, IPPP/19/42, MCNET-19-11},
}

@Article{Cacciari:2011ma,
  author        = {Cacciari, Matteo and Salam, Gavin P. and Soyez, Gregory},
  journal       = {Eur. Phys. J. C},
  title         = {{FastJet user manual}},
  year          = {2012},
  pages         = {1896},
  volume        = {72},
  archiveprefix = {arXiv},
  doi           = {10.1140/epjc/s10052-012-1896-2},
  eprint        = {1111.6097},
  primaryclass  = {hep-ph},
  reportnumber  = {CERN-PH-TH-2011-297},
}

@Article{Andersen:2010zx,
  author        = {Andersen, Jeppe R. and Arnold, Ken and Zeppenfeld, Dieter},
  journal       = {JHEP},
  title         = {{Azimuthal angle correlations for Higgs boson plus multi-jet events}},
  year          = {2010},
  pages         = {091},
  volume        = {06},
  archiveprefix = {arXiv},
  doi           = {10.1007/JHEP06(2010)091},
  eprint        = {1001.3822},
  primaryclass  = {hep-ph},
  reportnumber  = {CERN-PH-TH-2010-003, KA-TP-05-2010},
}

@Article{LHC,
  author  = {Evans, Lyndon and Bryant, Philip},
  journal = {JINST},
  title   = {{LHC Machine}},
  year    = {2008},
  pages   = {S08001},
  volume  = {3},
  doi     = {10.1088/1748-0221/3/08/S08001},
}

@Article{Davier:1992nw,
  author       = {Davier, M. and Duflot, L. and Le Diberder, F. and Rouge, A.},
  journal      = {Phys. Lett. B},
  title        = {{The Optimal method for the measurement of tau polarization}},
  year         = {1993},
  pages        = {411--417},
  volume       = {306},
  doi          = {10.1016/0370-2693(93)90101-M},
  reportnumber = {LAL-92-73, X-LPNHE-92-22},
}

@Article{Atwood:1991ka,
  author    = {Atwood, D. and Soni, A.},
  journal   = {Phys. Rev. D},
  title     = {Analysis for magnetic moment and electric dipole moment form factors of the top quark via ${\mathit{e}}^{+}$${\mathit{e}}^{\mathrm{\ensuremath{-}}}$\ensuremath{\rightarrow}tt\ifmmode\bar\else\textasciimacron\fi{}},
  year      = {1992},
  month     = {04},
  pages     = {2405--2413},
  volume    = {45},
  doi       = {10.1103/PhysRevD.45.2405},
  issue     = {7},
  numpages  = {0},
  publisher = {American Physical Society},
  url       = {https://link.aps.org/doi/10.1103/PhysRevD.45.2405},
}

@Article{Diehl:1993br,
  author       = {Diehl, M. and Nachtmann, O.},
  journal      = {Z. Phys. C},
  title        = {{Optimal observables for the measurement of three gauge boson couplings in $e^+ e^- \rightarrow W^+ W^-$}},
  year         = {1994},
  pages        = {397--412},
  volume       = {62},
  doi          = {10.1007/BF01555899},
  reportnumber = {HD-THEP-93-37},
}

@Article{ATLAS:2020rej,
  author        = {{ATLAS Collaboration}},
  journal       = {Eur. Phys. J. C},
  title         = {{Higgs boson production cross-section measurements and their EFT interpretation in the $4\ell $ decay channel at $\sqrt{s}=$13 TeV with the ATLAS detector}},
  year          = {2020},
  note          = {[Erratum: Eur. Phys. J. C 81 (2021) 29, Erratum: Eur. Phys. J. C 81 (2021) 398]},
  number        = {10},
  pages         = {957},
  volume        = {80},
  archiveprefix = {arXiv},
  collaboration = {ATLAS},
  doi           = {10.1140/epjc/s10052-020-8227-9},
  eprint        = {2004.03447},
  primaryclass  = {hep-ex},
  reportnumber  = {CERN-EP-2020-034},
}

@Article{ALEPH:2004klc,
  author        = {{ALEPH Collaboration}},
  journal       = {Phys. Lett. B},
  title         = {{Improved measurement of the triple gauge-boson couplings $\gamma$WW and ZWW in $e^{+}e^{-}$ collisions}},
  year          = {2005},
  pages         = {7--26},
  volume        = {614},
  collaboration = {ALEPH},
  doi           = {10.1016/j.physletb.2005.03.058},
  reportnumber  = {CERN-PH-EP-2004-065},
}

@Article{OPAL:2000wbs,
  author        = {{OPAL Collaboration}},
  journal       = {Eur. Phys. J. C},
  title         = {{Measurement of $W$ boson polarizations and CP violating triple gauge couplings from $W^{+} W^{-}$ production at LEP}},
  year          = {2001},
  pages         = {229--240},
  volume        = {19},
  archiveprefix = {arXiv},
  collaboration = {OPAL},
  doi           = {10.1007/s100520100602},
  eprint        = {hep-ex/0009021},
  reportnumber  = {CERN-EP-2000-113},
}

@Article{Hagiwara:1993ck,
  author       = {Hagiwara, Kaoru and Ishihara, S. and Szalapski, R. and Zeppenfeld, D.},
  journal      = {Phys. Rev. D},
  title        = {{Low energy effects of new interactions in the electroweak boson sector}},
  year         = {1993},
  pages        = {2182--2203},
  volume       = {48},
  doi          = {10.1103/PhysRevD.48.2182},
  reportnumber = {MAD-PH-737, UT-635, KEK-TH-356, KEK-PREPRINT-92-214},
}

@Article{Brivio:2017btx,
  author        = {Brivio, Ilaria and Jiang, Yun and Trott, Michael},
  journal       = {JHEP},
  title         = {{The SMEFTsim package, theory and tools}},
  year          = {2017},
  pages         = {070},
  volume        = {12},
  archiveprefix = {arXiv},
  doi           = {10.1007/JHEP12(2017)070},
  eprint        = {1709.06492},
  primaryclass  = {hep-ph},
}

@Article{Brivio:2020onw,
  author        = {Brivio, Ilaria},
  journal       = {JHEP},
  title         = {{SMEFTsim 3.0 \textemdash{} a practical guide}},
  year          = {2021},
  pages         = {073},
  volume        = {04},
  archiveprefix = {arXiv},
  doi           = {10.1007/JHEP04(2021)073},
  eprint        = {2012.11343},
  primaryclass  = {hep-ph},
}

@Article{Brehmer:2017lrt,
  author        = {Brehmer, Johann and Kling, Felix and Plehn, Tilman and Tait, Tim M. P.},
  journal       = {Phys. Rev. D},
  title         = {{Better Higgs-CP tests through information geometry}},
  year          = {2018},
  number        = {9},
  pages         = {095017},
  volume        = {97},
  archiveprefix = {arXiv},
  doi           = {10.1103/PhysRevD.97.095017},
  eprint        = {1712.02350},
  primaryclass  = {hep-ph},
  reportnumber  = {UCI-HEP-TR-2017-14},
}

@Misc{SupplementalMaterial,
  note = {See Supplemental Material at \url{http://link.aps.org/supplemental/10.1103/PhysRevLett.131.061802} for BDT input variable distributions and categorization selections.},
}

@Misc{atlasdetector,
  note = { ATLAS uses a right-handed coordinate system with its origin at the nominal interaction point (IP) in the center of the detector and the $z$-axis along the beam pipe. The $x$-axis points from the IP to the center of the LHC ring, and the $y$-axis points upwards. Cylindrical coordinates ($r, \phi$) are used in the transverse plane, $\phi$ being the azimuthal angle around the z-axis. The pseudorapidity is defined in terms of the polar angle $\theta$ as $\eta$ = $\mathrm{-ln~ tan(\theta/2)}$. Angular distance is measured in units of $\Delta R \equiv \sqrt{(\Delta \eta)^2+(\Delta \phi)^2}$.},
}

@Article{Cacciari:2008gp,
     author    = "Cacciari, Matteo and Salam, Gavin P. and Soyez, Gregory",
     title     = "{The anti-\(k_{t}\) jet clustering algorithm}",
     journal   = "JHEP",
     volume    = "04",
     year      = "2008",
     pages     = "063",
     eprint    = "0802.1189",
     archivePrefix = "arXiv",
     primaryClass  =  "hep-ph",
     doi       = "10.1088/1126-6708/2008/04/063",
     SLACcitation  = "%%CITATION = 0802.1189;%%"
}

@Article{Butterworth:2015oua,
      author         = "Butterworth, Jon and others",
      title          = "{PDF4LHC recommendations for LHC Run II}",
      journal        = "J. Phys. G",
      volume         = "43",
      year           = "2016",
      pages          = "023001",
      doi            = "10.1088/0954-3899/43/2/023001",
      eprint         = "1510.03865",
      archivePrefix  = "arXiv",
      primaryClass   = "hep-ph",
      reportNumber   = "OUTP-15-17P, SMU-HEP-15-12, TIF-UNIMI-2015-14,
                        LCTS-2015-27, CERN-PH-TH-2015-249",
      SLACcitation   = "%%CITATION = ARXIV:1510.03865;%%"
}

@article{Sjostrand:2014zea,
      author         = "Sj{\"o}strand, Torbj{\"o}rn and Ask, Stefan and Christiansen,
                        Jesper R. and Corke, Richard and Desai, Nishita and Ilten,
                        Philip and Mrenna, Stephen and Prestel, Stefan and
                        Rasmussen, Christine O. and Skands, Peter Z.",
      title          = "{An introduction to PYTHIA 8.2}",
      journal        = "Comput. Phys. Commun.",
      volume         = "191",
      year           = "2015",
      pages          = "159",
      doi            = "10.1016/j.cpc.2015.01.024",
      eprint         = "1410.3012",
      archivePrefix  = "arXiv",
      primaryClass   = "hep-ph",
      reportNumber   = "LU-TP-14-36, MCNET-14-22, CERN-PH-TH-2014-190,
                        FERMILAB-PUB-14-316-CD, DESY-14-178, SLAC-PUB-16122,
                        --FERMILAB-PUB-14-316-CD",
      SLACcitation   = "%%CITATION = ARXIV:1410.3012;%%"
}

@Article{Alwall:2014hca,
      author         = "Alwall, J. and Frederix, R. and Frixione, S. and Hirschi,
                        V. and Maltoni, F. and Mattelaer, O. and Shao, H. -S. and
                        Stelzer, T. and Torrielli, P. and Zaro, M.",
      title          = "{The automated computation of tree-level and
                        next-to-leading order differential cross sections, and
                        their matching to parton shower simulations}",
      journal        = "JHEP",
      volume         = "07",
      year           = "2014",
      pages          = "079",
      doi            = "10.1007/JHEP07(2014)079",
      eprint         = "1405.0301",
      archivePrefix  = "arXiv",
      primaryClass   = "hep-ph",
      reportNumber   = "CERN-PH-TH-2014-064, CP3-14-18, LPN14-066, MCNET-14-09,
                        ZU-TH-14-14",
      SLACcitation   = "%%CITATION = ARXIV:1405.0301;%%"
}

@article{Frederix:2012ps,
      author         = "Frederix, Rikkert and Frixione, Stefano",
      title          = "{Merging meets matching in MC@NLO}",
      journal        = "JHEP",
      volume         = "12",
      year           = "2012",
      pages          = "061",
      doi            = "10.1007/JHEP12(2012)061",
      eprint         = "1209.6215",
      archivePrefix  = "arXiv",
      primaryClass   = "hep-ph",
      reportNumber   = "CERN-PH-TH-2012-247, ZU-TH-21-12",
      SLACcitation   = "%%CITATION = ARXIV:1209.6215;%%"
}

@article{Hamilton:2012rf,
      author         = "Hamilton, Keith and Nason, Paolo and Oleari, Carlo and
                        Zanderighi, Giulia",
      title          = "{Merging H/W/Z + 0 and 1 jet at NLO with no merging
                        scale: a path to parton shower + NNLO matching}",
      journal        = "JHEP",
      volume         = "05",
      year           = "2013",
      pages          = "082",
      doi            = "10.1007/JHEP05(2013)082",
      eprint         = "1212.4504",
      archivePrefix  = "arXiv",
      primaryClass   = "hep-ph",
      reportNumber   = "CERN-PH-TH-2012-356",
      SLACcitation   = "%%CITATION = ARXIV:1212.4504;%%"
}

@article{Hamilton:2012np,
      author         = "Hamilton, Keith and Nason, Paolo and Zanderighi, Giulia",
      title          = "{MINLO: multi-scale improved NLO}",
      journal        = "JHEP",
      volume         = "10",
      year           = "2012",
      pages          = "155",
      doi            = "10.1007/JHEP10(2012)155",
      eprint         = "1206.3572",
      archivePrefix  = "arXiv",
      primaryClass   = "hep-ph",
      reportNumber   = "CERN-PH-TH-2012-166, OUTP-12-11P, MCNET-12-07",
      SLACcitation   = "%%CITATION = ARXIV:1206.3572;%%"
}

@article{deFlorian:2016spz,
      author         = "de Florian, D. and others",
      title          = "{Handbook of LHC Higgs Cross Sections: 4. Deciphering the
                        Nature of the Higgs Sector}",
      collaboration  = "LHC Higgs Cross Section Working Group",
      doi            = "10.23731/CYRM-2017-002",
      year           = "2016",
      eprint         = "1610.07922",
      archivePrefix  = "arXiv",
      primaryClass   = "hep-ph",
      reportNumber   = "FERMILAB-FN-1025-T, CERN-2017-002-M",
      SLACcitation   = "%%CITATION = ARXIV:1610.07922;%%"
}

@article{Hamilton:2013fea,
      author         = "Hamilton, Keith and Nason, Paolo and Re, Emanuele and
                        Zanderighi, Giulia",
      title          = "{NNLOPS simulation of Higgs boson production}",
      journal        = "JHEP",
      volume         = "10",
      year           = "2013",
      pages          = "222",
      doi            = "10.1007/JHEP10(2013)222",
      eprint         = "1309.0017",
      archivePrefix  = "arXiv",
      primaryClass   = "hep-ph",
      reportNumber   = "MCNET-13-11, CERN-PH-TH-2013-205, OUTP-13-18P",
      SLACcitation   = "%%CITATION = ARXIV:1309.0017;%%"
}

@article{Hamilton:2015nsa,
      author         = "Hamilton, Keith and Nason, Paolo and Zanderighi, Giulia",
      title          = "{Finite quark-mass effects in the NNLOPS POWHEG+MiNLO
                        Higgs generator}",
      journal        = "JHEP",
      volume         = "05",
      year           = "2015",
      pages          = "140",
      doi            = "10.1007/JHEP05(2015)140",
      eprint         = "1501.04637",
      archivePrefix  = "arXiv",
      primaryClass   = "hep-ph",
      reportNumber   = "CERN-PH-TH-2015-006",
      SLACcitation   = "%%CITATION = ARXIV:1501.04637;%%"
}

@article{Campbell:2012am,
      author         = "Campbell, John M. and Ellis, R. Keith and Frederix,
                        Rikkert and Nason, Paolo and Oleari, Carlo and Williams,
                        Ciaran",
      title          = "{NLO Higgs boson production plus one and two jets using
                        the POWHEG BOX, MadGraph4 and MCFM}",
      journal        = "JHEP",
      volume         = "07",
      year           = "2012",
      pages          = "092",
      doi            = "10.1007/JHEP07(2012)092",
      eprint         = "1202.5475",
      archivePrefix  = "arXiv",
      primaryClass   = "hep-ph",
      reportNumber   = "FERMILAB-PUB-12-040-T, CERN-PH-TH-2012-048",
      SLACcitation   = "%%CITATION = ARXIV:1202.5475;%%"
}

@article{Catani:2007vq,
      author         = "Catani, Stefano and Grazzini, Massimiliano",
      title          = "{Next-to-Next-to-Leading-Order Subtraction Formalism in Hadron Collisions and
                        its Application to Higgs-boson Production at the Large Hadron Collider}",
      journal        = "Phys. Rev. Lett.",
      volume         = "98",
      year           = "2007",
      pages          = "222002",
      doi            = "10.1103/PhysRevLett.98.222002",
      eprint         = "hep-ph/0703012",
      archivePrefix  = "arXiv",
      primaryClass   = "hep-ph",
      SLACcitation   = "%%CITATION = HEP-PH/0703012;%%"
}

@article{Nason:2009ai,
      author         = "Nason, Paolo and Oleari, Carlo",
      title          = "{NLO Higgs boson production via vector-boson fusion
                        matched with shower in POWHEG}",
      journal        = "JHEP",
      volume         = "02",
      year           = "2010",
      pages          = "037",
      doi            = "10.1007/JHEP02(2010)037",
      eprint         = "0911.5299",
      archivePrefix  = "arXiv",
      primaryClass   = "hep-ph",
      SLACcitation   = "%%CITATION = ARXIV:0911.5299;%%"
}

@article{Anastasiou:2016cez,
      author         = "Anastasiou, Charalampos and Duhr, Claude and Dulat, Falko
                        and Furlan, Elisabetta and Gehrmann, Thomas and Herzog,
                        Franz and Lazopoulos, Achilleas and Mistlberger, Bernhard",
      title          = "{High precision determination of the gluon fusion Higgs
                        boson cross-section at the LHC}",
      journal        = "JHEP",
      volume         = "05",
      year           = "2016",
      pages          = "058",
      doi            = "10.1007/JHEP05(2016)058",
      eprint         = "1602.00695",
      archivePrefix  = "arXiv",
      primaryClass   = "hep-ph",
      reportNumber   = "CP3-16-01, ZU-TH-27-15, NIKHEF-2016-004,
                        CERN-TH-2016-006",
      SLACcitation   = "%%CITATION = ARXIV:1602.00695;%%"
}

@article{Anastasiou:2015ema,
      author         = "Anastasiou, Charalampos and Duhr, Claude and Dulat, Falko
                        and Herzog, Franz and Mistlberger, Bernhard",
      title          = "{Higgs Boson Gluon-Fusion Production in QCD at Three
                        Loops}",
      journal        = "Phys. Rev. Lett.",
      volume         = "114",
      year           = "2015",
      pages          = "212001",
      doi            = "10.1103/PhysRevLett.114.212001",
      eprint         = "1503.06056",
      archivePrefix  = "arXiv",
      primaryClass   = "hep-ph",
      reportNumber   = "CERN-PH-TH-2015-055, CP3-15-07",
      SLACcitation   = "%%CITATION = ARXIV:1503.06056;%%"
}

@article{Dulat:2018rbf,
      author         = "Dulat, Falko and Lazopoulos, Achilleas and Mistlberger,
                        Bernhard",
      title          = "{iHixs 2 -- Inclusive Higgs cross sections}",
      journal        = "Comput. Phys. Commun.",
      volume         = "233",
      year           = "2018",
      pages          = "243-260",
      doi            = "10.1016/j.cpc.2018.06.025",
      eprint         = "1802.00827",
      archivePrefix  = "arXiv",
      primaryClass   = "hep-ph",
      reportNumber   = "CERN-TH-2018-019, SLAC-PUB-17222",
      SLACcitation   = "%%CITATION = ARXIV:1802.00827;%%"
}

@article{Harlander:2009mq,
      author         = "Harlander, Robert V. and Ozeren, Kemal J.",
      title          = "{Finite top mass effects for hadronic Higgs production at
                        next-to-next-to-leading order}",
      journal        = "JHEP",
      volume         = "11",
      year           = "2009",
      pages          = "088",
      doi            = "10.1088/1126-6708/2009/11/088",
      eprint         = "0909.3420",
      archivePrefix  = "arXiv",
      primaryClass   = "hep-ph",
      reportNumber   = "WUB-09-10",
      SLACcitation   = "%%CITATION = ARXIV:0909.3420;%%"
}

@article{Harlander:2009bw,
      author         = "Harlander, Robert V. and Ozeren, Kemal J.",
      title          = "{Top mass effects in Higgs production at
                        next-to-next-to-leading order QCD: Virtual corrections}",
      journal        = "Phys. Lett. B",
      volume         = "679",
      year           = "2009",
      pages          = "467-472",
      doi            = "10.1016/j.physletb.2009.08.012",
      eprint         = "0907.2997",
      archivePrefix  = "arXiv",
      primaryClass   = "hep-ph",
      reportNumber   = "WUB-09-08",
      SLACcitation   = "%%CITATION = ARXIV:0907.2997;%%"
}

@article{Harlander:2009my,
      author         = "Harlander, Robert V. and Mantler, Hendrik and Marzani,
                        Simone and Ozeren, Kemal J.",
      title          = "{Higgs production in gluon fusion at
                        next-to-next-to-leading order QCD for finite top mass}",
      journal        = "Eur. Phys. J. C",
      volume         = "66",
      year           = "2010",
      pages          = "359-372",
      doi            = "10.1140/epjc/s10052-010-1258-x",
      eprint         = "0912.2104",
      archivePrefix  = "arXiv",
      primaryClass   = "hep-ph",
      reportNumber   = "WUB-09-18, MAN-HEP-2009-44",
      SLACcitation   = "%%CITATION = ARXIV:0912.2104;%%"
}

@article{Pak:2009dg,
      author         = "Pak, Alexey and Rogal, Mikhail and Steinhauser, Matthias",
      title          = "{Finite top quark mass effects in NNLO Higgs boson
                        production at LHC}",
      journal        = "JHEP",
      volume         = "02",
      year           = "2010",
      pages          = "025",
      doi            = "10.1007/JHEP02(2010)025",
      eprint         = "0911.4662",
      archivePrefix  = "arXiv",
      primaryClass   = "hep-ph",
      reportNumber   = "SFB-CPP-09-116, TTP09-43",
      SLACcitation   = "%%CITATION = ARXIV:0911.4662;%%"
}

@article{Actis:2008ug,
      author         = "Actis, Stefano and Passarino, Giampiero and Sturm,
                        Christian and Uccirati, Sandro",
      title          = "{NLO electroweak corrections to Higgs boson production at
                        hadron colliders}",
      journal        = "Phys. Lett. B",
      volume         = "670",
      year           = "2008",
      pages          = "12-17",
      doi            = "10.1016/j.physletb.2008.10.018",
      eprint         = "0809.1301",
      archivePrefix  = "arXiv",
      primaryClass   = "hep-ph",
      reportNumber   = "PITHA-08-20, SFB-CPP-08-62, TTP08-38",
      SLACcitation   = "%%CITATION = ARXIV:0809.1301;%%"
}

@article{Actis:2008ts,
      author         = "Actis, Stefano and Passarino, Giampiero and Sturm,
                        Christian and Uccirati, Sandro",
      title          = "{NNLO computational techniques: The cases \(H \to \gamma\gamma\) and \(H \to gg\)}",
      journal        = "Nucl. Phys. B",
      volume         = "811",
      year           = "2009",
      pages          = "182-273",
      doi            = "10.1016/j.nuclphysb.2008.11.024",
      eprint         = "0809.3667",
      archivePrefix  = "arXiv",
      primaryClass   = "hep-ph",
      reportNumber   = "PITHA-08-24, SFB-CPP-08-73, TTP08-42",
      SLACcitation   = "%%CITATION = ARXIV:0809.3667;%%"
}

@article{Bonetti:2018ukf,
      author         = "Bonetti, Marco and Melnikov, Kirill and Tancredi,
                        Lorenzo",
      title          = "{Higher order corrections to mixed QCD-EW contributions
                        to Higgs boson production in gluon fusion}",
      journal        = "Phys. Rev. D",
      volume         = "97",
      year           = "2018",
      number         = "5",
      pages          = "056017",
      doi            = "10.1103/PhysRevD.97.056017",
      eprint         = "1801.10403",
      archivePrefix  = "arXiv",
      primaryClass   = "hep-ph",
      reportNumber   = "CERN-TH-2018-011, TTP18-004",
      SLACcitation   = "%%CITATION = ARXIV:1801.10403;%%",
      related = "Bonetti:2018ukf-err",
      relatedstring  = "Erratum:",
}

@article{Ciccolini:2007jr,
     author         = "Ciccolini, M. and Denner, Ansgar and Dittmaier, S.",
     title          = "{Strong and Electroweak Corrections to the Production of a Higgs Boson + 2 Jets via Weak Interactions at the Large Hadron Collider}",
     journal        = "Phys. Rev. Lett.",
     volume         = "99",
     pages          = "161803",
     doi            = "10.1103/PhysRevLett.99.161803",
     year           = "2007",
     eprint         = "0707.0381",
     archivePrefix  = "arXiv",
     primaryClass   = "hep-ph",
}

@Article{Ciccolini:2007ec,
    author    = "Ciccolini, Mariano and Denner, Ansgar and Dittmaier, Stefan",
    title     = "{Electroweak and QCD corrections to Higgs production via
                 vector-boson fusion at the CERN LHC}",
    journal   = "Phys. Rev. D",
    volume    = "77",
    year      = "2008",
    pages     = "013002",
    eprint    = "0710.4749",
    archivePrefix = "arXiv",
    primaryClass  =  "hep-ph",
    doi       = "10.1103/PhysRevD.77.013002",
    SLACcitation  = "%%CITATION = 0710.4749;%%"
}

@Article{Bolzoni:2010xr,
    author    = "Bolzoni, Paolo and Maltoni, Fabio and Moch, Sven-Olaf and
                 Zaro, Marco",
    title     = "{Higgs Boson Production via Vector-Boson Fusion at Next-to-Next-to-Leading Order in QCD}",
    journal   = "Phys. Rev. Lett.",
    volume    = "105",
    year      = "2010",
    pages     = "011801",
    eprint    = "1003.4451",
    archivePrefix = "arXiv",
    primaryClass  =  "hep-ph",
    doi       = "10.1103/PhysRevLett.105.011801",
    SLACcitation  = "%%CITATION = 1003.4451;%%"
}

@article{Denner:2014cla,
      author         = "Denner, Ansgar and Dittmaier, Stefan and Kallweit, Stefan
                        and M{\"u}ck, Alexander",
      title          = "{HAWK 2.0: A Monte Carlo program for Higgs production
                        in vector-boson fusion and Higgs strahlung at hadron
                        colliders}",
      journal        = "Comput. Phys. Commun.",
      volume         = "195",
      year           = "2015",
      pages          = "161-171",
      doi            = "10.1016/j.cpc.2015.04.021",
      eprint         = "1412.5390",
      archivePrefix  = "arXiv",
      primaryClass   = "hep-ph",
      reportNumber   = "FR-PHENO-2014-013, MITP-14-101, TTK-14-36",
      SLACcitation   = "%%CITATION = ARXIV:1412.5390;%%"
}

@Booklet{TMVA,
      author         = "Hoecker, Andreas and others",
      title          = "{TMVA - Toolkit for Multivariate Data Analysis}",
      year           = "2007",
      eprint         = "physics/0703039",
      archivePrefix  = "arXiv",
      primaryClass   = "physics.data-an",
}

@Article{PERF-2007-01,
    author         = "{ATLAS Collaboration}",
    title          = "{The ATLAS Experiment at the CERN Large Hadron Collider}",
    journal        = "JINST",
    volume         = "3",
    year           = "2008",
    pages          = "S08003",
    doi            = "10.1088/1748-0221/3/08/S08003",
    primaryClass   = "hep-ex",
}

@Article{SOFT-2010-01,
    author         = "{ATLAS Collaboration}",
    title          = "{The ATLAS Simulation Infrastructure}",
    journal        = "Eur. Phys. J. C",
    volume         = "70",
    year           = "2010",
    pages          = "823",
    doi            = "10.1140/epjc/s10052-010-1429-9",
    eprint         = "1005.4568",
    archivePrefix  = "arXiv",
    primaryClass   = "physics.ins-det",
}

@Article{STDM-2011-05,
    author         = "{ATLAS Collaboration}",
    title          = "{Measurement of the isolated diphoton cross section in \(pp\) collisions at \(\sqrt{s} = 7\,\text{TeV}\) with the ATLAS detector}",
    journal        = "Phys. Rev. D",
    volume         = "85",
    year           = "2012",
    pages          = "012003",
    doi            = "10.1103/PhysRevD.85.012003",
    reportNumber   = "CERN-PH-EP-2011-088",
    eprint         = "1107.0581",
    archivePrefix  = "arXiv",
    primaryClass   = "hep-ex",
}

@Article{HIGG-2012-27,
    author         = "{ATLAS Collaboration}",
    title          = "{Observation of a new particle in the search for the Standard Model Higgs boson with the ATLAS detector at the LHC}",
    journal        = "Phys. Lett. B",
    volume         = "716",
    year           = "2012",
    pages          = "1",
    doi            = "10.1016/j.physletb.2012.08.020",
    reportNumber   = "CERN-PH-EP-2012-218",
    eprint         = "1207.7214",
    archivePrefix  = "arXiv",
    primaryClass   = "hep-ex",
}

@Article{HIGG-2013-02,
    author         = "{ATLAS Collaboration}",
    title          = "{Measurements of Higgs boson production and couplings in diboson final states with the ATLAS detector at the LHC}",
    journal        = "Phys. Lett. B",
    volume         = "726",
    year           = "2013",
    pages          = "88",
    doi            = "10.1016/j.physletb.2013.08.010",
    reportNumber   = "CERN-PH-EP-2013-103",
    eprint         = "1307.1427",
    archivePrefix  = "arXiv",
    primaryClass   = "hep-ex",
    related        = "HIGG-2013-02-err",
    relatedstring  = "Erratum:",
}

@Article{HIGG-2013-08,
    author         = "{ATLAS Collaboration}",
    title          = "{Measurement of Higgs boson production in the diphoton decay channel in \(pp\) collisions at center-of-mass energies of \(7\) and \(8\,\text{TeV}\) with the ATLAS detector}",
    journal        = "Phys. Rev. D",
    volume         = "90",
    year           = "2014",
    pages          = "112015",
    doi            = "10.1103/PhysRevD.90.112015",
    reportNumber   = "CERN-PH-EP-2014-198",
    eprint         = "1408.7084",
    archivePrefix  = "arXiv",
    primaryClass   = "hep-ex",
}

@Article{PERF-2014-03,
    author         = "{ATLAS Collaboration}",
    title          = "{Performance of pile-up mitigation techniques for jets in \(pp\) collisions at \(\sqrt{s} = 8\,\text{TeV}\) using the ATLAS detector}",
    journal        = "Eur. Phys. J. C",
    volume         = "76",
    year           = "2016",
    pages          = "581",
    doi            = "10.1140/epjc/s10052-016-4395-z",
    reportNumber   = "CERN-PH-EP-2015-206",
    eprint         = "1510.03823",
    archivePrefix  = "arXiv",
    primaryClass   = "hep-ex",
}

@Article{PERF-2016-06,
    author         = "{ATLAS Collaboration}",
    title          = "{Identification and rejection of pile-up jets at high pseudorapidity with the ATLAS detector}",
    journal        = "Eur. Phys. J. C",
    volume         = "77",
    year           = "2017",
    pages          = "580",
    doi            = "10.1140/epjc/s10052-017-5081-5",
    reportNumber   = "CERN-EP-2017-055",
    eprint         = "1705.02211",
    archivePrefix  = "arXiv",
    primaryClass   = "hep-ex",
    related        = "PERF-2016-06-err",
    relatedstring  = "Erratum:",
}

@Article{TRIG-2016-01,
    author         = "{ATLAS Collaboration}",
    title          = "{Performance of the ATLAS trigger system in 2015}",
    journal        = "Eur. Phys. J. C",
    volume         = "77",
    year           = "2017",
    pages          = "317",
    doi            = "10.1140/epjc/s10052-017-4852-3",
    reportNumber   = "CERN-EP-2016-241",
    eprint         = "1611.09661",
    archivePrefix  = "arXiv",
    primaryClass   = "hep-ex",
}

@Article{HIGG-2017-13,
    author         = "{ATLAS Collaboration}",
    title          = "{Constraints on Higgs boson properties using \(WW^{*}(\to e\nu\mu\nu) jj\) production in \(36.1\,\text{fb}^{-1}\) of \(\sqrt{s} = 13\,\text{TeV}\) \(pp\) collisions with the ATLAS detector}",
    journal        = "Eur. Phys. J. C",
    volume         = "82",
    year           = "2021",
    pages          = "622",
    doi            = "10.1140/epjc/s10052-022-10366-1",
    reportNumber   = "CERN-EP-2021-096",
    eprint         = "2109.13808",
    archivePrefix  = "arXiv",
    primaryClass   = "hep-ex",
}

@Article{PERF-2017-02,
    author         = "{ATLAS Collaboration}",
    title          = "{Measurement of the photon identification efficiencies with the ATLAS detector using LHC Run~2 data collected in 2015 and 2016}",
    journal        = "Eur. Phys. J. C",
    volume         = "79",
    year           = "2019",
    pages          = "205",
    doi            = "10.1140/epjc/s10052-019-6650-6",
    reportNumber   = "CERN-EP-2018-216",
    eprint         = "1810.05087",
    archivePrefix  = "arXiv",
    primaryClass   = "hep-ex",
}

@Article{EGAM-2018-01,
    author         = "{ATLAS Collaboration}",
    title          = "{Electron and photon performance measurements with the ATLAS detector using the 2015--2017 LHC proton--proton collision data}",
    journal        = "JINST",
    volume         = "14",
    year           = "2019",
    pages          = "P12006",
    doi            = "10.1088/1748-0221/14/12/P12006",
    reportNumber   = "CERN-EP-2019-145",
    eprint         = "1908.00005",
    archivePrefix  = "arXiv",
    primaryClass   = "hep-ex",
}

@Article{HIGG-2018-14,
    author         = "{ATLAS Collaboration}",
    title          = "{Test of CP invariance in vector-boson fusion production of the Higgs boson in the \(H\to\tau\tau\) channel in proton--proton collisions at \(\sqrt{s} = 13\,\text{TeV}\) with the ATLAS detector}",
    journal        = "Phys. Lett. B",
    volume         = "805",
    year           = "2020",
    pages          = "135426",
    doi            = "10.1016/j.physletb.2020.135426",
    reportNumber   = "CERN-EP-2020-009",
    eprint         = "2002.05315",
    archivePrefix  = "arXiv",
    primaryClass   = "hep-ex",
}

@Article{JETM-2018-05,
    author         = "{ATLAS Collaboration}",
    title          = "{Jet energy scale and resolution measured in proton--proton collisions at \(\sqrt{s} = 13\,\text{TeV}\) with the ATLAS detector}",
    journal        = "Eur. Phys. J. C",
    volume         = "81",
    year           = "2020",
    pages          = "689",
    doi            = "10.1140/epjc/s10052-021-09402-3",
    reportNumber   = "CERN-EP-2020-083",
    eprint         = "2007.02645",
    archivePrefix  = "arXiv",
    primaryClass   = "hep-ex",
}

@Article{JETM-2018-06,
    author         = "{ATLAS Collaboration}",
    title          = "{Optimisation of large-radius jet reconstruction for the ATLAS detector in \(13\,\text{TeV}\) proton--proton collisions}",
    journal        = "Eur. Phys. J. C",
    volume         = "81",
    year           = "2020",
    pages          = "334",
    doi            = "10.1140/epjc/s10052-021-09054-3",
    reportNumber   = "CERN-EP-2020-134",
    eprint         = "2009.04986",
    archivePrefix  = "arXiv",
    primaryClass   = "hep-ex",
}

@Article{PIX-2018-001,
    author         = "Abbott, B. and others",
    title          = "{Production and integration of the ATLAS Insertable B-Layer}",
    journal        = "JINST",
    volume         = "13",
    year           = "2018",
    pages          = "T05008",
    doi            = "10.1088/1748-0221/13/05/T05008",
    eprint         = "1803.00844",
    archivePrefix  = "arXiv",
    primaryClass   = "physics.ins-det",
}

@Article{TRIG-2018-05,
    author         = "{ATLAS Collaboration}",
    title          = "{Performance of electron and photon triggers in ATLAS during LHC Run~2}",
    journal        = "Eur. Phys. J. C",
    volume         = "80",
    year           = "2020",
    pages          = "47",
    doi            = "10.1140/epjc/s10052-019-7500-2",
    reportNumber   = "CERN-EP-2019-169",
    eprint         = "1909.00761",
    archivePrefix  = "arXiv",
    primaryClass   = "hep-ex",
}

@Article{HIGG-2020-08,
    author         = "{ATLAS Collaboration}",
    title          = "{Study of the CP property of the Higgs boson to electroweak boson coupling in the VBF \(H\to\gamma\gamma\) channel with the ATLAS detector}",
    year           = "2022",
    reportNumber   = "CERN-EP-2022-134",
    eprint         = "2208.02338",
    archivePrefix  = "arXiv",
    primaryClass   = "hep-ex",
}
\clearpage
 
\begin{flushleft}
\hypersetup{urlcolor=black}
{\Large The ATLAS Collaboration}

\bigskip

\AtlasOrcid[0000-0002-6665-4934]{G.~Aad}$^\textrm{\scriptsize 101}$,
\AtlasOrcid[0000-0002-5888-2734]{B.~Abbott}$^\textrm{\scriptsize 119}$,
\AtlasOrcid[0000-0002-7248-3203]{D.C.~Abbott}$^\textrm{\scriptsize 102}$,
\AtlasOrcid[0000-0002-1002-1652]{K.~Abeling}$^\textrm{\scriptsize 55}$,
\AtlasOrcid[0000-0002-8496-9294]{S.H.~Abidi}$^\textrm{\scriptsize 29}$,
\AtlasOrcid[0000-0002-9987-2292]{A.~Aboulhorma}$^\textrm{\scriptsize 35e}$,
\AtlasOrcid[0000-0001-5329-6640]{H.~Abramowicz}$^\textrm{\scriptsize 150}$,
\AtlasOrcid[0000-0002-1599-2896]{H.~Abreu}$^\textrm{\scriptsize 149}$,
\AtlasOrcid[0000-0003-0403-3697]{Y.~Abulaiti}$^\textrm{\scriptsize 116}$,
\AtlasOrcid[0000-0003-0762-7204]{A.C.~Abusleme~Hoffman}$^\textrm{\scriptsize 136a}$
% ...
\bigskip
\\

% ...
$^{29}$Physics Department, Brookhaven National Laboratory, Upton NY; United States of America.\\
% ...
$^{35}$$^{(a)}$Facult\'e des Sciences Ain Chock, R\'eseau Universitaire de Physique des Hautes Energies - Universit\'e Hassan II, Casablanca;$^{(b)}$Facult\'{e} des Sciences, Universit\'{e} Ibn-Tofail, K\'{e}nitra;$^{(c)}$Facult\'e des Sciences Semlalia, Universit\'e Cadi Ayyad, LPHEA-Marrakech;$^{(d)}$LPMR, Facult\'e des Sciences, Universit\'e Mohamed Premier, Oujda;$^{(e)}$Facult\'e des sciences, Universit\'e Mohammed V, Rabat;$^{(f)}$Institute of Applied Physics, Mohammed VI Polytechnic University, Ben Guerir; Morocco.\\
% ...
$^{55}$II. Physikalisches Institut, Georg-August-Universit\"{a}t G\"ottingen, G\"ottingen; Germany.\\
% ...
$^{101}$CPPM, Aix-Marseille Universit\'e, CNRS/IN2P3, Marseille; France.\\
$^{102}$Department of Physics, University of Massachusetts, Amherst MA; United States of America.\\
% ...
$^{116}$Department of Physics, New York University, New York NY; United States of America.\\
% ...
$^{119}$Homer L. Dodge Department of Physics and Astronomy, University of Oklahoma, Norman OK; United States of America.\\
% ...
$^{136}$$^{(a)}$Departamento de F\'isica, Pontificia Universidad Cat\'olica de Chile, Santiago;$^{(b)}$Millennium Institute for Subatomic physics at high energy frontier (SAPHIR), Santiago;$^{(c)}$Instituto de Investigaci\'on Multidisciplinario en Ciencia y Tecnolog\'ia, y Departamento de F\'isica, Universidad de La Serena;$^{(d)}$Universidad Andres Bello, Department of Physics, Santiago;$^{(e)}$Instituto de Alta Investigaci\'on, Universidad de Tarapac\'a, Arica;$^{(f)}$Departamento de F\'isica, Universidad T\'ecnica Federico Santa Mar\'ia, Valpara\'iso; Chile.\\
% ...
$^{149}$Department of Physics, Technion, Israel Institute of Technology, Haifa; Israel.\\
$^{150}$Raymond and Beverly Sackler School of Physics and Astronomy, Tel Aviv University, Tel Aviv; Israel.\\
% ...
$^{a}$ Also Affiliated with an institute covered by a cooperation agreement with CERN.\\
$^{b}$ Also at An-Najah National University, Nablus; Palestine.\\
$^{c}$ Also at Borough of Manhattan Community College, City University of New York, New York NY; United States of America.\\
$^{d}$ Also at Bruno Kessler Foundation, Trento; Italy.\\
$^{e}$ Also at Center for High Energy Physics, Peking University; China.\\
$^{f}$ Also at Centro Studi e Ricerche Enrico Fermi; Italy.\\
% ...

\end{flushleft}


%

\clearpage

\clearpage

\end{document}